
\pubnum{177}
\date={April, 1993}
\titlepage
\vskip1.0cm
\title{\bf The Stability of the Minisuperspace}
\vskip2.0cm
\centerline{Atushi Ishikawa\footnote*{
        e-mail: 402g0028@ex.ecip.osaka-u.ac.jp}
		 and Toshiki Isse\footnote{**}{
        e-mail: 402g0027@ex.ecip.osaka-u.ac.jp}}

\vskip1.5cm

\centerline {\it  Department of Physics}
\centerline {\it Osaka University, Toyonaka, Osaka 560, Japan}

\vskip0.5cm

\vskip3cm

\abstract{The stability of the minisuperspace model of the early universe
is studied by solving the Wheeler-DeWitt
equation numerically. We consider a system of
Einstein gravity with a scalar field. When we solve the Wheeler-DeWitt
equation, we pick up some inhomogeneous wave modes from the infinite number of
modes adequately: degrees of freedom of the superspace are
restricted to a finite one. We show that the minisuperspace is stable when
a scale factor ($a$) of the universe is larger than a few times of the
Planck
length, while it becomes unstable when $a$ is comparable to the Planck length.}

\medskip

\vfill

\eject

\endpage


\REF\Haw{S.~W.~Hawking, {\it Nucl.~Phys.} {\bf B239} (1984) 257.}

\REF\Hal{J.~J.~Halliwell and S.~W.~Hawking, {\it Phys.~Rev.}
{\bf D31} (1985) 1777.}

\REF\Mis{C.W.~Misner, in {\it Magic Without Magic}, edited by J.~Klauder
(Freeman, Sanfrancisco, 1972);
C.~W.~Misner, {\it Phys.~Rev.} {\bf 186} (1969) 1319.}

\REF\DeW{B.~S.~DeWitt, {\it Phys.~Rev.} {\bf 160} (1967) 1113.}

\REF\Rya{M.~Ryan, {\it Hamiltonian Cosmology} (Spring, Berlin, 1972).}

\REF\Sin{S.~Sinha and B.L.~Hu, {\it Phys.~Rev.}
{\bf D44} (1991) 1028.}

\REF\Kuc{K.V.~Kuchar and M.P.~Ryan,~Jr., {\it Phys.~Rev.}
{\bf D40} (1990) 3982.}

\REF\Har{J.~B.~Hartle and S.~W.~Hawking, {\it Phys.~Rev.}
{\bf D28} (1983) 2960.}

\REF\Vil{A.~Vilenkin, {\it ibid} {\bf 33} (1986) 3560.}

\REF\Lif{E.M.~Lifschiz and I.M.~Khalatnikov, {\it Adv.~Phys.}
{\bf 12} (1963) 185.}

\REF\Ger{U.H.~Gerlach and U.K.~Sengupta, {\it Phys.~Rev.}
{\bf D18} (1978) 1773.}

\REF\Lou{J.J.~Halliwell and J.~Louko, {\it Phys.~Rev.}
{\bf D39} (1989) 2206.}

\REF\Louu{J.J.~Halliwell and J.~Louko, {\it Phys.~Rev.}
{\bf D42} (1990) 3997.}

\REF\Gar{L.J.~Garay, J.J.~Halliwell and G.A.M.~Marugan, {\it Phys.~Rev.}
{\bf D43} (1991) 2572.}

\REF\Wada{S.~Wada, {\it Phys.~Rev.} {\bf D34} (1986) 2272.}


\def\int{\intop\nolimits}


\chapter {Introduction}

interaction
Hamiltonian


 When we discuss the classical dynamics of the Universe, it is very useful
and physically reasonable to assume that the Universe has certain
symmetries. A well-known example is the Friedmann Universe which has the
spatial homogeneity and isotropy.
gravitational (

In the past few decades, many people have discussed the quantum treatment
of the Universe by solving zero-energy Schrodinger equations. The
Schrodinger equations are decomposed into the Wheeler-DeWitt equation and
the momentum constraints which are the quantum versions of the Hamiltonian
and the momentum constraints respectively. A quantum state of the Universe
is described by a wave function $\Psi$ that is a function on the infinite
dimensional space called the superspace $W$. The superspace $W$ consists of
all three-geometry and matter field configuration on a three-surface $S$.
 The Wheeler-DeWitt and the momentum constraint equations are
infinite-dimensional partial differential equations on the superspace.
Apart from many difficult conceptual problems, it is too difficult to solve
such infinite-dimensional differential equations without any approximation.
It makes the problem tractable to reduce the degree of freedom of $W$ to
finite by assuming a given symmetry. Such finite dimensional approximation
to $W$ is called the minisuperspace approximation.

  The quantization of the Universe was first introduced by DeWitt
\refmark {\DeW} using the Friedmann minisuperspace approximation. The
homogeneous anisotropic cases were then studied by Misner \refmark {\Mis}
and
others \refmark {\Rya}. Moreover the minisuperspace model has received
renewed attention since the appearance of the paper of Hartle and Hawking
\refmark {\Har} on proposals for no-boundary boundary conditions of $ \Psi $.
Although the boundary conditions can be formally stated in general cases,
analytical calculations are performed only in the minisuperspace model
\refmark {\Lou, \Louu, \Gar} and all inhomogeneous modes are omitted
before quantization.
Some people's attempts to introduce the inhomogeneous modes along the
path-integral method using WKB approximation are given in \refmark {\Hal,
\Wada}.
Above naive setting of the non-zero modes equal to zero before quantization
apparently violates the uncertainty principle and therefore a question
arises; whether or not the minisuperspace approximation is a physically
reasonable approximation of some true quantum gravity.
Although this problem has been well known to many people working in the
field of quantum cosmology for a long time, only a few people
\refmark {\Kuc, \Sin} have checked the validity of the minisuperspace model.
In Ref.[{\Kuc}] Kuchar and Ryan discussed the quantum cosmology of a Taub
universe embedded in a mixmaster universe to assess the validity of the
minisuperspace model. They analyzed the lower-dimensional homogeneous
superspace model (Taub universe) and the higher-(but finite)dimensional
homogeneous superspace model( mixmaster universe) in a region where exact
solutions for both models exist. In their work the inhomogeneous modes are
not considered.
  While in Ref.[{\Sin}] Sinha and Hu analyzed an effect of discarding an
infinite number of inhomogeneous modes using WKB approximation. However,
they found that the WKB approximation is of limited validity and that it is
important to go beyond it when classical background is not available and
full quantum behavior of the minisuperspace sector is important.

 In this paper we examine the stability of the Friedmann minisuperspace
model by numerically solving the Wheeler-DeWitt equation. We consider a
system of Einstein gravity with a scalar field as a model of the quantum
cosmology.
In order to check the stability of the minisuperspace model, we have only
to consider a configuration space of the metric and the scalar fields near
the minisuperspace. In other words, we can regard all inhomogeneous modes
as very small. In this approximation the inhomogeneous modes are decoupled
to one another. When we solve the Wheeler-DeWitt equation, we pick up some
inhomogeneous wave modes from infinite number of the other modes
adequately; a degree of freedom of the superspace of this system is
restricted to finite. Our method is therefore similar to that of Kuchar and
Ryan. We consider some higher-dimensional (but finite) superspace which
includes the lower-dimensional Friedmann minisuperspace.
  It should be noted that, as we compute wave functions of the Universe
numerically, our analysis is not necessarily limited within the validity
region in which WKB approximation is applicable.



 The present paper is organized as follows. In Sec.II we review the Hamiltonian
formalism of Einstein gravity with a scalar field and its canonical
treatment of the quantization. In Sec.III we show the work on a homogeneous
and isotropic minisuperspace model, which is known as the Friedmann
model, with a scalar field. We extend this model to all the
matter and gravitational degree of freedom in Sec.IV, following the
work of Halliwell and Hawking work \refmark {\Hal}. In Sec.V we pick up some
particular inhomogeneous modes from the perturbed minisuperspace's
infinite wave modes and solve the Wheeler-DeWitt equation on the newly
defined superspace numerically.
Finally, in Sec.VI we summarize the paper and conclude.


 \chapter {Canonical formulation and quantization}

 We consider a system of Einstein gravity and a scalar field. The action is
$$
I=\int {(L_g+L_m)},
\eqn\ee
$$
where
$$
\eqalignno{
\eqname\aa
&L_g={{m_p^2} \over {16\pi }}\sqrt {-g}R, &\aa\cr
& \cr
\eqname\bb
&L_m={1 \over 2}(-g^{\mu \nu }\partial _\mu \Phi \partial _\nu \Phi
                   -m^2\Phi ^2). &\bb \cr}
$$
 The (1+3) decomposition of the metric reads
$$
ds^2=-(N^2-N^iN_i)dt^2+2N_idx^idt+h_{ij}dx^idx^j,
\eqn\ee
$$
where $N$,$N^i$ and $h_{ij}$ are the lapse, shift functions
and the spatial metric.
The spatial indices $i$ runs from 1 to 3.
The canonical conjugate momentum of $h_{ij}$
and that of $\Phi$ are respectively given by
$$
\pi ^{ij}={{\partial L_g} \over {\partial \dot h_{ij}}}=-{{\sqrt hm_p^2}
\over {16\pi }}(K^{ij}-h^{ij}K)
\eqn\ee
$$
and
$$
\pi _\Phi ={{\partial L_m} \over {\partial \dot \Phi }}=N^{-1}\sqrt h(\dot
\Phi -N^i{{\partial \Phi } \over {\partial x^i}}),
\eqn\ee
$$
where $K_{ij}$ and $K$ are the extrinsic curvature and its trace part
respectively. The Hamiltonian is obtained as
$$
\eqalign{H&=\int {(\pi ^{ij}\dot h_{ij}}+\pi _\Phi \dot \Phi -L_g-L_m)dx^3\cr
  &=\int {(NH_0}+N^iH_i)dx^3,\cr}
\eqn\ee
$$
where
$$
\eqalignno{
\eqname\cc
H_0=& 16\pi m_p^{-2}G_{ijkl}\pi ^{ij}\pi ^{kl}
           -{{\sqrt hm_p^2} \over {16\pi }}{}^{(3)}R \cr
   &+{{\sqrt h} \over 2}({{\pi _\Phi ^2} \over h}
		   +h^{ij}\partial _i\Phi \partial _j\Phi +m^2\Phi ^2), &\cc\cr
\eqname\dd
H^i=&-2\pi ^{ij}_{\left| j \right.}
                 +h^{ij}\partial _j\Phi \pi _\Phi &\dd
  \cr}
$$
and
$$
G_{ijkl}={1 \over 2}h^{-1/2}(h_{ik}h_{jl}+h_{il}h_{jk}-h_{ij}h_{kl}).
\eqn\ee
$$

We follow the Dirac's prescription of quantization for the constrained system.
The momentum are replaced by the differential operators
$$
\pi ^{ij}(x)\rightarrow -i{\delta  \over {\delta h_{ij}(x)}}
\eqn\ee
$$
and
$$
\pi _\Phi (x)\rightarrow -i{\delta  \over {\delta \Phi (x)}}.
\eqn\ee
$$
The Wheeler-DeWitt equation and the momentum constraints are given as
$$
H_0\Psi =0
\eqn\ee
$$
and
$$
H^i\Psi =0,
\eqn\ee
$$
where the momentum in $H_0$ and $H_i$ are substituted
by the differential operators.


\chapter {The Friedmann Minisuperspace Model}

In this section we briefly review the Friedmann minisuperspace model.
It is assumed that a metric of this model has spatial homogeneity and
isotropy, which has the advantage of reducing the infinite dimensional
superspace to a finite dimensional minisuperspace. The metric of this
model is given as
$$
ds^2=\sigma ^2(-N^2dt^2+a^2d\Omega _3^2),
\eqn\ee
$$
where $d\Omega _3^2$ is a metric of a unit three-sphere and $\sigma$ is a
normalization
factor which has been introduced for convenience $\sigma ^2=2/3\pi m_p^2$.
The action of this model is described by
$$
I=-{1 \over 2}\int {dtNa^3}({{\dot a^2} \over {N^2a^2}}-{1 \over
{a^2}}-{{\dot \phi ^2} \over {N^2}}+m^2\phi ^2).
\eqn\ee
$$
It should be noted that the action of this model contains a rescaled scalar
field $\phi /\sqrt 2\pi \sigma$
and mass $m/\sigma $ and that the scalar field is constant on
the hypersurface of constant $t$. The classical Hamiltonian is
$$
H={1 \over 2}N(-a^{-1}\pi _a^2+a^{-3}\pi _\phi ^2-a+a^3m^2\phi ^2),
\eqn\ee
$$
where
$$
\pi _a=-{{a\dot a} \over N}
\eqn\ee
$$
and
$$
\pi _\phi ={{a^3\dot \phi } \over N}.
\eqn\ee
$$
Because the action has a time reparametrization invariance, the classical
field equations are constrained by a Hamiltonian constraint equation $H=0$.
The Wheeler-DeWitt equation is
 $$
{N \over 2}e^{-3\alpha }({{\partial ^2} \over {\partial \alpha
^2}}-{{\partial ^2} \over {\partial \phi ^2}}+e^{6\alpha }m^2\phi
^2-e^{-4\alpha })\Psi =0,
\eqn\ee
$$
where $\alpha =ln(a)$. One can regard eq.(3.6) as the hyperbolic differential
equation on the two dimensional minisuperspace with coordinates
($\alpha$, $\phi$).  We can therefore regard  $\alpha $ as a time coordinate.


\chapter {The Perturbed Friedmann Minisuperspace Model}

We first describe a metric near the Friedmann metric as
$$
ds^2=\sigma ^2\bigl\{-(N^2-N^iN_i)dt^2+2N_idx^idt+h_{ij}dx^idx^j\bigr\}.
\eqn\ee
$$
The metric of the spacial hypersurface $h_{ij}$ has the form
$$
h_{ij}=a^2(\Omega _{ij}+\epsilon _{ij}),
\eqn\ee
$$
where $\epsilon_{ij}$ is a small quantity around the metric of
unit three-sphere $\Omega _{ij}$.
Following Halliwel and Hawking[\Hal],the quantities
 $\epsilon_{ij}$ is expanded in harmonics as
$$
\eqalign{\varepsilon _{ij}=& \sum\limits_{n,l,m}
   {[\sqrt 6a_{nlm}\Omega _{ij}Q_{lm}^n/3+\sqrt 6b_{nlm}(P_{ij})_{lm}^n
   +\sqrt 2c_{lmn}^o(S_{ij}^o)_{lm}^n}\cr
  &+\sqrt 2c_{lmn}^e(S_{ij}^e)_{lm}^n+\sqrt 2d_{lmn}^o(G_{ij}^o)_{lm}^n
   +\sqrt 2d_{lmn}^e(G_{ij}^e)_{lm}^n].\cr}
\eqn\ee
$$
The lapse, shift and the scalar fields can be also expanded by harmonics
$$
\eqalignno{
\eqname\ff
&N=N_0[1+6^{-1/2}\sum\limits_{n,l,m} {g_{nlm}Q_{lm}^n}], &\ff\cr
\eqname\gg
&N_i=a\sum\limits_{n,l,m}
     {[6^{-1/2}k_{nlm}(P_i)_{lm}^n+2^{1/2}j_{nlm}(S_i)_{lm}^n] }, &\gg\cr
\eqname\hh
&\Phi=\sigma ^{-1}[2^{-1/2}\pi \phi (t)
      +\sum\limits_{n,l,m} {f_{nlm}Q_{lm}^n}].  &\hh\cr}
$$
The definition and the normalization of these harmonics are given in
the Appendix. The coefficients $a_{nlm}$,$b_{nlm}$,$c_{nlm}^o$,
$c_{nlm}^e$,$d_{nlm}^o$,$d_{nlm}^e$ ,$g_{nlm}$,$j_{nlm}$,$k_{nlm}$,$f_{nlm}$
are only functions of time.
We expand the action to all orders in the zero mode of
the fields ($a$, $\phi$ and $N_0$) but only to second order in the non-zero
modes ($a_{nlm}$, $b_{nlm}$,..., $f_{nlm}$);
$$
I=I_0(a,\phi ,N_0)+\sum\limits_n {I_n},
\eqn\ee
$$
where $I_0$ is the action of the Friedmann minisuperspace model and $I_n$
are terms quadratic in perturbations.  The exact forms of $I_n$ are given
as
$$
I_n=\int {dt(I_{gravity}^n}+I_{matter}^n),
\eqn\ee
$$
where
$$
\eqalign{I_{gravity}^n
  =& {e^{\alpha} \over 2}N_0
     \Bigl[(n^2-5/2)a_n^2/3
       +{{(n^2-7)(n^2-4)} \over {3(n^2-1)}}b_n^2-2(n^2-4)c_n^2 \cr
   &-(n^2+1)d_n^2+2(n^2-4)a_nb_n/3+2g_n
            \bigl \{ {(n^2-4)b_n+(n^2+1/2)a_n} \bigr \}/3\cr
   &+{1 \over {N_0^2}} \bigl \{ {-{{k_n^2} \over {3(n^2-1)}}
         +(n^2-4)j_n^2} \bigr \} \Bigr]\cr
   &+{{e^{3\alpha}} \over {2N_0}}
   \Bigl[-\dot a_n^2+{{(n^2-4)} \over {(n^2-1)}}\dot b_n^2
           +(n^2-4)\dot c_n^2+\dot d_n^2 \cr
   &+g_n \bigl \{ {2\dot \alpha \dot a_n
		        +{\dot \alpha}^2(3a_n-g_n)} \bigr \} \cr
   &+\dot \alpha \bigl \{ {-2a_n\dot a_n
       +8{{(n^2-4)} \over {(n^2-1)}}b_n\dot b_n
	   +8(n^2-4)c_n\dot c_n+8d_n\dot d_n} \bigr \} \cr
   &+{\dot \alpha}^2 \bigl \{{-3a_n^2/2+6{{(n^2-4)} \over {(n^2-1)}}b_n^2
                     +6(n^2-4)c_n^2+6d_n^2} \bigr \} \cr
   &+e^{-\alpha} \bigl \{ {k_n(-2\dot a_n/3
            +{{2(n^2-4)} \over {3(n^2-1)}}\dot b_n+2\dot \alpha g_n /3)
			-2(n^2-4)j_n\dot c_n} \bigr \}
	\Bigr]\cr}
\eqn\ee
$$
and
$$
\eqalign{I_{matter}^n
  =& {{N_0 e^{3 \alpha}} \over 2}
    \Bigl[{{(\dot f_n^2+6a_n\dot f_n\dot \phi )} \over {N_0^2}}
	        -m^2(f_n^2+6a_nf_n\phi )-e^{-2 \alpha}(n^2-1)f_n^2\cr
   &+{3 \over 2}({{\dot \phi ^2} \over {N_0^2}}-m^2\phi ^2)
     \bigl \{ {a_n^2-{{4(n^2-4)} \over {(n^2-1)}}b_n^2
          -4(n^2-4)c_n^2-4d_n^2} \bigr \} \cr
   &-g_n(2m^2f_n\phi +3m^2a_n\phi ^2+{{2\dot f_n\dot \phi } \over {N_0^2}}
                  +{{3a_n\dot \phi ^2} \over {N_0^2}})
				  -{{2k_nf_n\dot \phi } \over {e^{\alpha}N_0^2}}
	 \Bigr].\cr}
\eqn\ee
$$
One should note that we consider the configuration space near
the Friedmann minisuperspace and that the zero-mode of the fields
$a$,$\phi$ and $N_0$ are not classical solutions.
The Hamiltonian can be expressed as
$$
\eqalign{H=& N_0(H_{\left| 0 \right.}+\sum\limits_n {H_{\left| 2 \right.}^n}
           +\sum\limits_n {g_nH_{\left| 1 \right.}^n})\cr
           & +\sum\limits_n {(k_nH_{_1}^{(S)n}}+j_nH_{_1}^{(V)n}),\cr}
\eqn\ee
$$
where $H_{\left| 0 \right.}$ is the Hamiltonian of the Friedmann
minisuperspace model.
The second-order Hamiltonian $H_{\left| 2 \right.}$ is given as
$$
H_{\left| 2 \right.}=\sum\limits_n {H_{\left| 2 \right.}^n}=\sum\limits_n
{(H_{\left| 2 \right.}^{(S)n}}+H_{\left| 2 \right.}^{(V)n}+H_{\left| 2
\right.}^{(T)n}),
\eqn\ee
$$
where
$$
\eqalignno{
\eqname\ii
H_{\left| 2 \right.}^{(S)n}
  =&{1 \over {2 e^{3 \alpha}}}
  \Bigl[\bigl \{a_n^2/2
          +{{10(n^2-4)} \over {(n^2-1)}}b_n^2\bigr \} \pi _\alpha ^2
		  +\bigl \{15a_n^2/2
		  +{{6(n^2-4)} \over {(n^2-1)}}b_n^2 \bigr \} \pi _\phi ^2 \cr
   &-\pi _{a_n}^2+{{(n^2-1)} \over {(n^2-4)}}\pi _{b_n}^2
    +\pi _{f_n}^2+2a_n\pi _{a_n}\pi _\alpha
	+8b_n\pi _{b_n}\pi _\alpha \cr
   &-6 a_n\pi _{f_n}\pi _\phi
     -e^{4\alpha} \bigl \{{{(n^2-5/2)} \over 3}a_n^2
	 +{{(n^2-7)(n^2-4)} \over {3(n^2-1)}}b_n^2 \cr
   &+{{2(n^2-4)} \over 3} a_n b_n - (n^2-1)f_n^2 \bigr \} \cr
   &+e^{6\alpha} m^2 (f_n^2+6a_nf_n\phi )
    +e^{6 \alpha} m^2 \phi ^2
      \bigl\{ 3a_n^2/2-{{6(n^2-4)} \over {(n^2-1)}}b_n^2 \bigr \}
  \Bigr], &\ii \cr
\eqname\jj
H_{\left| 2 \right.}^{(V)n}
   =&{1 \over {2 e^{3 \alpha}}}
    \Bigl[ (n^2-4) c_n^2 (10\pi _\alpha ^2 + 6\pi _\phi ^2)
	       +\pi _{c_n}^2/(n^2-4) \cr
    &+8c_n \pi _{c_n} \pi _\alpha
	 +(n^2-4) c_n^2 (2e^{4\alpha}-6e^{6 \alpha} m^2 \phi ^2)
    \Bigr], &\jj \cr
\eqname\kk
H_{\left| 2 \right.}^{(T)n}
     =&{1 \over {2e^{3\alpha}}}
	   \Bigl[ d_n^2(10\pi _\alpha ^2 +6\pi _\phi ^2)
	       +\pi _{d_n}^2 +8d_n\pi _{d_n}\pi _\alpha \cr
      &+d_n^2 \bigl \{(n^2+1)e^{4 \alpha} -6e^{6 \alpha} m^2\phi ^2
	          \bigr \}
       \Bigr]. &\kk \cr}
$$
The first order Hamiltonian is
$$
\eqalign{H_{\left| 1 \right.}^n
  =& {1 \over {2e^{3 \alpha}}}
     \Bigl[ -a_n(\pi _\alpha ^2+3\pi _\phi ^2)
	        +2(\pi _{f_n}\pi _\phi -\pi _{a_n}\pi _\alpha ) \cr
   &+e^{6 \alpha} m^2 (3a_n\phi ^2+2f_n\phi )
    -{{2e^{4 \alpha}} \over 3} \bigl \{ (n^2-4)b_n+(n^2+2/1)a_n \bigr \}
     \Bigr].\cr}
\eqn\ee
$$
The shift parts of Hamiltonian are
$$
\eqalignno{
\eqname\ll
H_{_1}^{(S)n}=& {1 \over {3e^{3 \alpha}}}
               \Bigl[ -\pi _{a_n}+\pi _{b_n}
			   +\bigl \{a_n+{{4(n^2-4)} \over {(n^2-1)}}b_n
			    \bigr \} \pi _\alpha +3f_n\pi _\phi
			   \Bigr], &\ll \cr
\eqname\mm
H_{_1}^{(V)n}=& e^{-\alpha} \Bigl[\pi _{c_n}+4(n^2-4)c_n\pi _\alpha
                          \Bigr]. &\mm \cr}
$$
The Wheeler-DeWitt equation is decomposed into two parts,
$$
(H_{\left| 0 \right.}+\sum\limits_n {H_{\left| 2 \right.}^n})\Psi =0
\eqn\ee
$$
and
$$
H_{\left| 1 \right.}^n\Psi =0.
\eqn\ee
$$
Following Haliiwell and Hawking \refmark {\Hal}, we call eq(4.19)
the master equation
and eq(4.20) the linear Hamitonian constraint. The momentum constraints are
$$
H_{_1}^{(S)n}\Psi =0,
\eqn\ee
$$
$$H_{_1}^{(V)n}\Psi =0.
\eqn\ee
$$

\chapter {The Analysis of the Stability of the Minisuperspace}

Although we consider only the second order in the non-zero
modes, the master equation is difficult to solve technically.
To make the problem tractable we solve the master eq(4.19) numerically
by picking up some imhomogeneous wave modes from the infinite number of modes
adequately. In other words
we restrict the degree of freedom of the supersapce of this system
to a finite one.
The reason why we can pick up some particular inhomogeneous modes
is that the perturbation modes
are not coupled directly and that the master eq(4.19)
is expressed as a sum of the each mode.
There are many ways of picking up the inhomogeneous modes.
For example, we can fix the mode number $n$ for some value $n_0$.
Furthermore we can pick up only the scalar harmonics part of
the gravity ($a_n, b_n, g_n$
and $k_n$), the vector harmonics part ($c_n,j_n$) or the tensor part ($d_n$).
The easiest non-trivial one is described
in what follows.
For the metric,
we set the scalar type perturbations
($a_n$, $b_n$, $g_n$ and
$k_n$) equal to zero and consider only
the vector and the tensor type perturbations
($c_n, d_n$ and $j_n$) for some fixed $n$.
For the matter field, we also fix the mode number to $n$.
We consider a higher-dimensional (but finite) space $W'$
which includes the lower dimensional minisuperspace and then
we solve a reduced Wheeler-DeWitt equation on the space $W'$.
We are not suffered from the problem of solving the Wheeler-DeWitt equation
and the linear Hamiltonian constraint equation simultaneously,
because we omit the scalar type perturbations
($a_n, b_n, g_n$ and $k_n$).
We do not discuss the effect of the scalar harmonics part.

Following the above restriction,
we get a master equation on the restricted space $W'$:
$$
(H_{\left| 0 \right.}+ H_{\left| 2 \right.}^{(V)n}+H_{\left| 2
\right.}^{(T)n})\Psi =0.
\eqn
\ee
$$
Hereafter we call the eqation eq(5.1) the fixed master equation.
There are neither the scalar momentum constraint eq(4.21)
nor the linear Hamiltonian constraint eq(4.20)
because of the lack of $k_n$ and $g_n$.
Following Halliwell and Hawking \refmark {\Hal}, we can eliminate $c_n$ from
eq(5.1) by using the vector momentum constraint.
We can use the momentum constraint eq(4.22) to substitute
for the partial derivatives with respect to $c_n$
and then solve the resultant differential equation on $c_n=0$.
After we know the wave function on $c_n=0$, we can use the vector momentum
constraint eq(4.22) to calculate the wave function at other values of
$c_n$.

The momenta are replaced by the differential operators
$$
\pi _{\alpha} \rightarrow -i{\partial \over {\partial \alpha}},
\eqn
\ee
$$
$$
\pi _{d_n} \rightarrow -i{\partial \over {\partial d_n}},
\eqn
\ee
$$
$$
\pi _{\phi} \rightarrow -i{\partial \over {\partial \phi}}
\eqn
\ee
$$
and
$$
\pi _{f_n} \rightarrow -i{\partial \over {\partial f_n}}.
\eqn
\ee
$$
Consequently we get the modified fixed master equation:
$$
\eqalign{&
{N \over 2} {1 \over a^3} [(1-10 d_n^2)
(a^2 {{\partial ^2} \over {\partial a^2}}
+a {{\partial} \over {\partial a}})
-(1+6 d_n^2) {{\partial ^2} \over {\partial \phi ^2}}
-{{\partial ^2} \over {\partial d_n^2}}\cr
&-8 d_n a {{\partial ^2} \over {\partial d_n \partial a}}
-{{\partial ^2} \over {\partial f_n^2}}
+m^2 a^6 \phi^2 -a^4 \cr
&+d_n^2 \left({ (n^2 +1)a^4-6m^2 a^6 \phi^2} \right)
+(n^2 -1) a^4 f_n^2+m^2 a^6 f_n^2
] \Psi = 0. \cr}
\eqn
\ee
$$
This is the hyperbolic partial differential
equation. We can therefore regard the scale factor $a$ as time.

To analyze the stability of the minisuperspace, we first set a wave packet
which is centered around a minisuperspace sector as the initial condition,
$$
\Psi(a=a_0) \propto exp(-u_0 d_n^2),
\eqn
\ee
$$
where $u_0$ is an arbitrary constant Figs.(1,2).
We then investigate the development of the wave packet we prepared.
We impose the boundary condition of the wave function: $\Psi \rightarrow 0$
as the inhomogeneous perturbation $d_n$ or $f_n$ goes to infinity.
The free parameters of this system are the mass of the scalar field $m$,
the mode number $n$ and the initial value of the scale factor $a_0$.
We compute the development of the wave packet with respect to
the scale factor $a$
in the various cases of these parameters.
And the results are summarized in Tables.(1,2).
For the minisuperspace case, we can see that the wave function of the universe
grows exponentially, see Fig.(3).
For the perturbed minisuperspace case, if the wave packet
develops exponentially without much spreading into the surrounding
minisuperspace, we can consider the minisuperspce stable
(for example, in a case $a_0=20, m=0$ and $n=2$, see Fig.(1)).
On the other hand, if the wave packet dose not grow exponentially
or much spreading into the surrounding minisuperspace, we can say
the minisuperspace is unstable
(for example, in a case $a_0=2, m=0$ and $n=2$, see Fig.(2)).
Furthermore we compute the development of a standard deviation ($\Delta d_n$)
of the wave packet to check the minisuperspace is realy stable.
The standard deviation $\Delta d_n$ is used as an index which represents
the width of $\Psi$ in the direction of $d_n$.
The results are in the Fig.(4).
{}From this figure, we can conclude that
the minisuperspace is stable when the initial value of the scale
factor ($a_0$) of the universe is larger than a few times of the Planck
length . In this case, we can regard the wave packet (which grows
exponentially) approximately as the wave function of the universe
on the minisuperspace.
On the other hand, the minisuperspace becomes unstable
when $a_0$ is near the Planck length .
The results do not change even if the mass of the scalar field
or the mode number $n$ is set large, see Table.(1,2).


\chapter {Conclusion}

We examined the stability of the Friedmann minisuperspace model by
numerically solving the Wheeler-DeWitt equation. In our analysis a fixed
mode number $n$ is picked up and we consider only the scalar harmonics
part. We can conclude that the minisuperspace is stable when the initial
scale factor ($a_0$) of the Universe is larger than a few times of the
Plank length. The stability of the minisuperspace increases as $m$ or $n$
becomes large. While the minisuperspace is unstable when $a_0$ is about the
Plank length within the limitation of the numerical analysis ($n < 100,
m<10
and a_0 <100$).

\vskip2.5cm

Acknowledgments

We would like to acknowledge a careful reading of the manuscript by
Dr. H. Kunitomo.
We want to express our gratitude to other members High Energy physics group
in Osaka University for their warm encouragement.

\vskip2.5cm



\appendix

The scalar spherical harmonics $Q_{lm}^n(\chi ,\theta ,\varphi )$ are defined
as
$$
Q_{lm}^n(\chi ,\theta ,\varphi )=\Pi _l^n(\chi )Y_{lm}(\theta ,\varphi ),
\eqn\ee
$$
where $\Pi _l^n(\chi )$ are the Fock harmonics
\refmark {\Lif,\Ger} and $Y_{lm}(\theta ,\varphi )$ are the spherical
harmonics on $S^2$.
The scalar harmonics $Q_{lm}^n$ are the eigenfunctions
of the Laplacian operator on $S^3$,
$$
Q_{\   \left| k \right.}^{(n)\  \left| k \right.}=-(n^2-1)Q^{(n)},\  n=1,2,...
\eqn\ee
$$
The spherical harmonics $Q_{lm}^n$ constitute a complete orthogonal set for
the expansion of any scalar function on $S^3$.

The transverse vector harmonics $S_i^{(n)}(\chi ,\theta ,\varphi )$ are
the eigenfunctions of the Laplacian operator on $S^3$,
$$
\eqalign{&S_{i\   \left| k \right.}^{(n)\  \left| k
\right.}=-(n^2-2)S_i^{(n)},\  n=2,3,...\cr
  &S_i^{(n)\  \left| i \right.}=0.\cr}
\eqn\ee
$$
The exact expressions of $S_i^{(n)}$ are given in Ref[\Ger]
and $S_i^{(n)}$ are classified by as odd $(o)$ or even $(e)$ parts using
a parity transformation. The third vector harmonics $P_i$ are defined
by the scalar harmonics $Q^n_{lm}$,
$$
P_i={1 \over {(n^2-1)}}Q_{\left| i \right.},\  \ n=2,3,...
\eqn\ee
$$
and satisfy
$$
P_{i\   \left| k \right.}^{(n)\  \left| k \right.}
  =-(n^2-3)P_i^{(n)},\  P_i^{(n)\  \left| i \right.}=-Q.
\eqn\ee
$$
The three types of vector harmonics $S_i^{(o)}$,$S_i^{(e)}$ and $P_i$
constitute a complete orthogonal set for any vector function on $S^3$.

The transverse traceless tensor harmonics $G_{ij}^{(n)}$ are
the eigenfunctions of the Laplacian operators on $S^3$ which are
transverse and traceless,
$$
\eqalign{&G_{ij\   \left| k \right.}^{(n)\  \left| k \right.}
          =-(n^2-3)G_{ij}^{(n)},\  n=3,4,...\cr
         &G_{ij}^{(n)\  \left| i \right.}=0,\  G_i^{(n)i}=0.\cr}
\eqn\ee
$$
They are also classified as odd $G_{ij}^{(o)}$or even $G_{ij}^{(e)}$parts.
Explicit expressions for $G_{ij}^{(o)}$ and $G_{ij}^{(e)}$ are given
in Ref[\Ger]. The traceless tensor harmonics
$S^{(o)}_{ij}$ and $S^{(e)}_{ij}$ are defined, both for odd and even, by
$$
S_{ij}=S_{\left. i \right|j}+S_{\left. j \right|i }
\eqn\ee
$$
and they satisfy
$$
\eqalign{&S_{ij}^{\   \left| j \right.}=-(n^2-4)S_i,\cr
  &S_{ij}^{\   \left| {ij} \right.}=0,\cr
  &S_{ij\   \left| k \right.}^{\  \left| k \right.}=-(n^2-6)S_{ij}.\cr}
\eqn\ee
$$
The two tensors $Q_{ij}$ and $P_{ij}$ are defined by the scalar harmonics,
$$
\eqalign{&Q_{ij}={1 \over 3}\Omega _{ij}Q,\  n=1,2,...\cr
  &P_{ij}={1 \over {(n^2-1)}}Q_{\left| {ij} \right.}+{1 \over 3}\Omega
_{ij}Q,\  n=2,3,...\cr}
\eqn\ee
$$
and satisfy the following equations,
$$
\eqalign{&P_{ij}^{\   \left| j \right.}=-{2 \over 3}(n^2-4)P_i,\cr
  &P_{ij\   \left| k \right.}^{\  \left| k \right.}=-(n^2-7)P_{ij},\cr
  &P_{ij}^{\   \left| {ij} \right.}=-{2 \over 3}(n^2-4)Q.\cr}
\eqn\ee
$$
The six tensor harmonics $G_{ij}^{(o)}$,$G_{ij}^{(e)}$,$S^{(o)}_{ij}$,
$S^{(o)}_{ij}$,$Q_{ij}$ and $P_{ij}$ constitute a complete orthogonal set
for any symmetric second rank tensor function on $S^3$.

The normalization and the orthogonality relations are described
by the following equations,
$$
\eqalign{&\int {d\mu }Q_{lm}^nQ_{l'm'}^{n'}=\delta ^{nn'}\delta
_{mm'}\delta _{ll'},\cr
  &\int {d\mu }(P_i)_{lm}^n(P^i)_{l'm'}^{n'}={1 \over {n^2-1}}\delta
^{nn'}\delta _{mm'}\delta _{ll'},\cr
  &\int {d\mu }(P_{ij})_{lm}^n(P^{ij})_{l'm'}^{n'}={{2(n^2-4)} \over
{3(n^2-1)}}\delta ^{nn'}\delta _{mm'}\delta _{ll'},\cr
  &\int {d\mu }(S_{ij})_{lm}^n(S^{ij})_{l'm'}^{n'}=2(n^2-4)\delta
^{nn'}\delta _{mm'}\delta _{ll'},\cr
  &\int {d\mu }(G_{ij})_{lm}^n(G^{ij})_{l'm'}^{n'}=\delta ^{nn'}\delta
_{mm'}\delta _{ll'},\cr}
\eqn\ee
$$
where $d\mu$ is a integration measure on $S^3$,
$$
\eqalign{d\mu& =dx^3(\det \Omega _{ij})^{1/2}\cr
  &=\sin \theta d\chi d\theta d\varphi .\cr}
\eqn\ee
$$
The reader may find further details on the harmonics in Ref[\Lif] and [\Ger].


\refout


\end